\documentclass[12pt]{article}
\usepackage{amstex}
\usepackage{epic}
\usepackage{graphics}

\columnwidth\textwidth

\begin{document}

\newcommand{\be}{\begin{equation}}
\newcommand{\ee}{\end{equation}}
\newcommand{\beann}{\begin{eqnarray*}}
\newcommand{\eeann}{\end{eqnarray*}}
\newcommand{\bea}{\begin{eqnarray}}
\newcommand{\eea}{\end{eqnarray}}
\newcommand{\nn}{\nonumber}
\newcommand{\ben}{\begin{enumerate}}
\newcommand{\een}{\end{enumerate}}
\newtheorem{df}{Definition}
\newtheorem{thm}{Theorem}
\newtheorem{lem}{Lemma}
\newtheorem{prop}{Proposition}
\begin{titlepage}

\noindent
\hspace*{11cm} BUTP-01/24 \\
\vspace*{1cm}
\begin{center}
{\LARGE Pair of null gravitating shells I. \newline
         Space of solutions and its symmetries} 

\vspace{2cm}

P. H\'{a}j\'{\i}\v{c}ek and I. Kouletsis\\
Institute for Theoretical Physics \\
University of Bern \\
Sidlerstrasse 5, CH-3012 Bern, Switzerland \\
\vspace*{2cm}

December 2001 \\ \vspace*{1cm}

\nopagebreak[4]

\begin{abstract}
  The dynamical system constituted by two spherically symmetric thin shells
  and their own gravitational field is studied. The shells can be
  distinguished from each other, and they can intersect. At each intersection,
  they exchange energy on the Dray, 't~Hooft and Redmount formula. There are
  bound states: if the shells intersect, one, or both, external shells can be
  bound in the field of internal shells. The space of all solutions to
  classical dynamical equations has six components; each has the trivial
  topology but a non trivial boundary. Points within each component are
  labeled by four parameters. Three of the parameters determine the geometry
  of the corresponding solution spacetime and shell trajectories and the
  fourth describes the position of the system with respect to an observer
  frame. An account of symmetries associated with spacetime diffeomorphisms is
  given. The group is generated by an infinitesimal time shift, an
  infinitesimal dilatation and a time reversal.
\end{abstract}

\end{center}

\end{titlepage}

\section{Introduction}
\label{sec:intro}

A remarkable feature of gravitation is that of gravitational collapse. The
theory predicts spectacular stages of collapse such as the formation of black
holes and the inevitability of a final singularity if a black hole is formed.
There is some evidence that black holes indeed exist.  Alas, the singularity
means that the classical theory breaks down.

In this situation, it is natural to look to quantum theory for a remedy.
However, a quantum theory of gravity as a logically self-consistent theory
backed up by observation and experiment does not exist, and it seems unlikely
that it will exist as such a theory in the near future.  This provides our
motivation first for focusing on a particular, truly physical, problem such as
gravitational collapse and, second, for working with simplified models.  Such
models can give useful hints even before a full-fledged quantum gravity is
invented. One may even compare them to the theory of atomic spectra before the
invention of quantum electrodynamics.  By simplifying technicalities, they
enable us to construct exact quantum theories instead of just, say,
semi-classical approximations thereof. In fact, no semi-classical
approximation would do near the singularities.

Of course, if we reject semi-classical approximations as the leading idea then
we are in for trouble. First, some semi-classical approximation must indeed be
valid somewhere. At the very least this should be near the infinity, where the
fields and interactions are weak, but it may even be as far down as near the
black hole horizons of large astronomical objects, as current observations
seem to be bearing out. We must thus confirm that the constructed theory does
not have a wrong classical limit. Second, a semi-classical approach can always
be formulated as a theory on a fixed classical potential. For gravity, this
effectively boils down to a quantum theory on a fixed classical background
spacetime. That has a great advantage: the conceptual difficulties of quantum
gravity can be avoided. If we abandon semi-classical approximation then we are
exposed at once to these problems.

Our method to meet conceptual problems is based on a suitably chosen set of
gauge-invariant quantities. The quantum theory is then formulated exclusively
in terms of these quantities. Since the gauge group in question is the
diffeomorphism group of the whole spacetime, gauge invariance includes the
property of being an integral of motion; these quantities are called {\em
  Dirac observables}. The use of Dirac observables in quantum gravity goes
back to Bergmann \cite{bergmann}, who also articulated two difficulties
associated with them: First, there seemed to be no no-trivial
dynamics\footnote{``Frozen dynamics'' was Bergmann's term.} and, second, no
single Dirac observable was known for general relativity (in general). A third
difficulty has been discovered recently \cite{torre}: No Dirac observables
exist in general relativity that are local functionals of the fields.

However, if the physical problem at hand can be given the form of a {\em
  scattering} problem so that every relevant question can be answered by a
measurement in the asymptotically flat region of spacetime, then the above
difficulties can be solved. This has been outlined in \cite{H-I} for the case
of quantum field theory on curved spacetime. Indeed, the gauge group of
asymptotically flat spacetime does not include all diffeomorphisms but only
those that do not move points at infinity. The diffeomorphisms that move such
points are not gauge, but symmetry transformations. There is then enough
symmetry to construct a non-trivial dynamics of Dirac observables \cite{haji}.
Moreover, there are even two complete sets of Dirac observables, which are the
usual asymptotic in- and out-fields of scattering theory. For gravity, these
observables have been called ``asymptotic invariants'' by DeWitt \cite{DW}
within a perturbative scattering theory and ``radiative modes'' by Ashtekar
\cite{A} within his exact theory of ``asymptotic quantization''. Finally,
asymptotic fields are local in their corresponding asymptotic regions, and
there is no necessity to write them as (non-local) functionals of the fields
at finite times\footnote{Of course, the $S$-matrix is non local, as one could
  see if one managed to write it in $x$-representation, but this does not seem
  to cause problems.}.

In order to construct a quantum theory based on Dirac observables, Poisson
brackets within the chosen set of observables must be known: the {\em algebra
  of Dirac observables}. We shall calculate this algebra by specifying a
gauge. The method has been invented by Kucha\v{r} \cite{K} and fully developed
by H\'{a}j\'{\i}\v{c}ek and Kijowski \cite{H-Kij}. It is based on the
so-called {\em Kucha\v{r} decomposition}: all variables are split into Dirac
observables describing the physical degrees of freedom, the {\em embedding
  variables} that represent the gauge variables and the {\em embedding
  momenta} that represent the dependent variables. The notions of {\it
  background manifold} and of {\it covariant gauge fixing} (necessary for the
definition of embedding variables and for making calculations) are explained
in \cite{H-Kij}.

This program has been successfully completed for a simplified collapse model,
see \cite{H-Kie} and \cite{H}. It consists of a single spherically symmetric
shell of null dust surrounded by its own gravitational field. The constructed
quantum dynamics has been unitary; the shell wave packet has been sent from
infinity by some observers, it has collapsed, bounced, re-expanded and has
been finally captured by the same observers at infinity. For sufficiently high
energy of the packet, most of the packet has passed the Schwarzschild radius
back and forth. This does not lead to contradiction because the quantum
Schwarzschild radius has been {\em black} (black hole) when the shell has been
passing it inwards, and it has been {\em white} (white hole) when the shell
has been passing it outwards.

There have been, however, some questions. First, the shell has a zero radial
extension and the spacetime under it is Minkowski spacetime. The canonical
formalism can be cast in the form of a shell moving on Minkowski spacetime
\cite{B-H}. It is then very simple and in fact natural to arrange the bounce!
But what happens if the system has non-zero radial extension? Then the
dynamics cannot very naturally be reformulated as motion on Minkowski
spacetime. Can a unitary quantum mechanics still be constructed?

Second, what is the nature of the metric outside the shell? For a quantum
shell in a spherically symmetric case, the metric belongs to dependent
variables so it must be a quantum metric. However, the notion of such a
quantum metric is not gauge invariant \cite{H}. Even classically, what could
be the role, say, of an ``average'' metric at a point of the background
manifold, even if a gauge choice makes such notion well-defined? The problem
is that in order for a tensor field to be a metric it must have signature
$+2$. Since the signature is not a linear property, the sum of two metrics
need not be any metric at all. The analogous notions of an ``operator of
metric'' at a given point of the background manifold and of its expectation
value in some quantum state do not make much sense either; they are not useful
quantities for the description of the quantum field outside the shell. How can
we then obtain any physically meaningful information about the field outside
the shell?

The two questions provide our motivation for the study of two-shells: It may
be the simplest system that has a non-zero extension; the second shell moves
in the field of the first one, that is, classically, in a Schwarzschild
spacetime. It seems then that there is no possibility to reduce everything to
a motion on Minkowski spacetime. Hence, the first question can be studied.
Moreover, the second shell probes the field of the first one. In a spherically
symmetric case, this is even literally so because any shell influences only
the part of spacetime that lies outside of it. If we manage to construct a
quantum theory, we may read from the asymptotic state of the out-going second
shell information about the field of the first one.

As yet, however, no quantum theory has been constructed. Only the classical
part of the work has been done: some complete (phase space spanning) sets of
Dirac observables have been found and their Poisson brackets have been
calculated. The calculation requires several steps so we have split it into
three papers, numbered I, II and III. The derivation of the results is
described in such a way that it is easy to generalize it to any number of
shells, and the results themselves are formulated so that their extension to
any number of shell is trivial.

The present paper, I, describes and parametrizes the set of all solutions. It
turns out that two shells form a very interesting and much less trivial system
than a single shell. The reason is that two shells can intersect each other,
and they interact in a relatively complicated way at the intersection.
Physically, the interaction can be defined as Raychaudhuri's effect: positive
matter crossing a null hypersurface enlarges the convergence of its null
generators in dependence of the matter density. Thus, an out-going shell is
left less divergent and so its energy is smaller after the crossing, while the
in-going shell is made more convergent and so has a larger energy after the
crossing. Technically, everything is determined for the shells by the
requirement that the metric is continuous even at the crossing point. The
resulting condition in the spherically symmetric case has been written down by
Dray and 't~Hooft \cite{DtH} and by Redmount \cite{Red}. The interaction has a
contact character. It is interesting that, if the shells intersect, the
out-going shell can be left in a bound state after the crossing, while the
in-going shell can be bound before the crossing (``bound'' means that it does
not reach infinity). All this gives a lot of structure to the space of
solutions. An intriguing question (that we don't answer in these three papers)
is whether a true Hamiltonian can be written down that contains the
interaction at the crossing.

Finally, the present paper also specifies the symmetries of the two-shell
system. They are the diffeomorphisms that move points at infinity and are
compatible with the spherical symmetry of the model. They can help us to
construct the time evolution as shown in \cite{haji}.

The plan of the paper is as follows. Sec.~\ref{sec:solutions} describes the
construction of all possible solutions to Einstein equations that are
spherically symmetric, contain two null shells and no other sources of
gravity. The space of all such solutions consists of six disconnected
subspaces. Each subspace has the trivial topology but a non-trivial
boundary. Coordinates we choose to parametrize the solutions in different
subspaces can be separated into two groups: geometric and frame-position
parameters. The geometric parameters determine the metric and the shell
trajectories up to isometry. The frame-position parameters define the position
of the system with respect to an (asymptotic) observer frame.

Sec.~\ref{sec:gauge} presents an example of gauge choice for one of the
subspaces. The set of spacetimes is turned to a set of metric fields and shell
trajectories on a background manifold. In Sec.~\ref{sec:symm}, the symmetries
of the space of solutions are specified. They consist of time shifts, time
reversal and dilatations (re-scalings). The gauge defined in
Sec.~\ref{sec:gauge} is employed for the calculation of the action of the
symmetries on embedding variables; this action is important for the
construction of the dynamics as described in \cite{haji}.

\section{Space of solutions}
\label{sec:solutions}
Spherically symmetric solutions with two null shells are well-known, cf.\ 
\cite{BI}, \cite{DtH} and \cite{Red}. In this section, a standardized
construction of these solutions is described. The construction depends on
certain parameters. If some of these parameters change, the physical
properties of the solution also change. Such parameters can be chosen as
coordinates in the physical phase space. Some other parameters change but the
physical properties remain the same. Such parameters describe pure gauge. A
bijective association of all solutions with the values of four parameters
lying in certain spaces will be established.

All spacetime solutions can be constructed using the following rules. The
metric in any shell-free part of the spacetime is the Schwarzschild one
corresponding to some value $M$ of the Schwarzschild mass parameter. We assume
that the innermost part is flat ($M=0$) and contains a regular center so that
Cauchy hypersurfaces have the topology of ${\mathbf R}^3$. The shells form
hypersurface boundaries of these pieces of Schwarzschild spacetimes. The
hypersurfaces are light-like with respect to the metrics of their neighboring
spacetimes. There must be coordinates in a neighborhood of any shell point
such that the metric in the neighborhood is continuous \cite{BI}. This holds
even for the crossing points of two shells, see \cite{DtH} and \cite{Red}.
Using these rules, we obtain all solutions. The energy density of the shell
matter is assumed to be positive.

The two shells are physically distinguishable. Different shells are described
by a parameter $s = 1,2$. These indices represent the properties of the shell
material by which we can recognize them. One can also imagine that the first
shell is green and the second red. However, the shell dynamics is invariant
with respect to shell permutation.

The assumption on the topology of Cauchy hypersurfaces implies that there is
only one infinity, that is only one scri, $\mathcal I$. We assume that the
observers live there and define a particular reference frame. With respect to
these observers, each shell is either out- or in-going (irrespectively
of whether the shell reaches the $\mathcal I$ or not). This property of a
shell is described by the parameter $\eta$: $\eta = +1$ if the shell is
out-going and $\eta = -1$ otherwise. Since the shells are distinguishable 
we have to consider six cases:
\begin{description}
\item[A:] Both shells are in-going, $\eta_1 = \eta_2 = -1$, with the first
  shell on the left.
\item[A':] Both shells are in-going, $\eta_1 = \eta_2 = -1$, with the second
  shell on the left.
\item[B:] Both shells are out-going, $\eta_1 = \eta_2 = +1$, with the first
  shell on the left.
\item[B':] Both shells are out-going, $\eta_1 = \eta_2 = +1$, with the second
  shell on the left.
\item[C:] The first shell is in- and the second is out-going, $\eta_1 =
  -\eta_2 = -1$.
\item[C':] The first shell is out- and the second is in-going, $\eta_1 =
  -\eta_2 = +1$.
\end{description}

Our construction will now be outlined case by case.

\subsection{Parallel shells}
\begin{figure}[h]
\begin{center}
\begin{picture}(120,180)
 \put(104,144){${\mathcal I}^+$} \put(70,50){${\mathcal I}^-$}
 \put(30,75){${\mathcal M}_l$} \put(60,110){${\mathcal M}_m$}
 \put(82,128){${\mathcal M}_r$}
 \dashline[50]{5}(60,140)(80,160)
 \dashline[50]{5}(30,130)(50,150)
 \dottedline[$\star$]{9}(0,160)(80,160)
 \dottedline{3}(0,0)(0,160)
 \drawline(0,0)(120,120)(80,160)
 \thicklines
 \drawline(0,160)(80,80) \drawline(40,160)(100,100)
\end{picture}
\end{center}
\caption{Penrose diagram of case A. The star line is the singularity, the
  dotted line is the regular center, the thick lines are the shells and the
  dashed lines are the Schwarzschild horizons. Case B can be considered as the
  time reversal of case A.}\label{fig:A}
\end{figure}
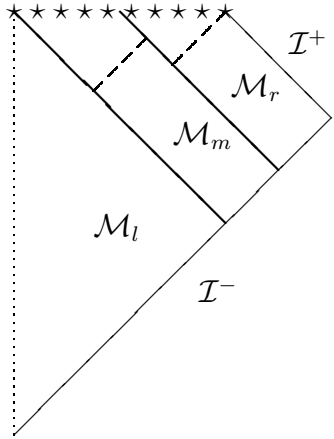

Cases A and A' are schematically depicted by Fig.~\ref{fig:A}; case A' differs
from A just by swapping the shells; we can deal with both cases simultaneously
by speaking about internal and external shells. The intershell spacetimes are
denoted by ${\mathcal M}_K$, $K = l,m,r$, where the subscripts stand for left,
middle and right.  The Schwarzschild mass of ${\mathcal M}_K$ is denoted by
$M_K$. The mass of ${\mathcal M}_l$ is zero.  $M_r$ is the total mass of the
system and $M_m$ is the total energy of the first shell.  The energy of the
second shell is $M_r - M_m$.

Any solution of this type with metric and shell trajectories can be
constructed as follows. Let $\overline{\mathcal M}_K$ be the maximal analytic
extension of the spacetime ${\mathcal M}_K$. The quotient $\overline{\mathcal
  M}_K/\text{SO}(3)$ of these spherically symmetric spacetimes are themselves
spacetimes, albeit two-dimensional. The spacetimes $\overline{\mathcal
  M}_K/\text{SO}(3)$ can be given time and space orientations. We assume that
the orientation of pasted spacetime agrees with the orientation of its parts.
 The notions of the out- and in-going shell can also be
defined more precisely: future oriented tangential vector is oriented in any
orthonormal two-bein to the right (left) for out- (in-)going shell. In this
way, the orientation of the shell motion is indeed well defined even if the
shell does not intersect infinity.

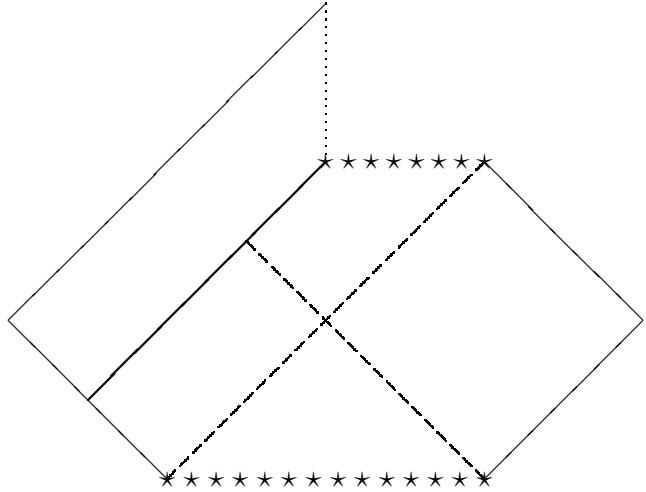
\begin{figure}[h]
\begin{center}
\begin{picture}(240,200)
 \dashline[50]{5}(60,0)(180,120)
 \dashline[50]{5}(90,90)(180,0)
 \dottedline[$\star$]{9}(60,0)(180,0)
 \dottedline[$\star$]{9}(120,120)(180,120)
 \dottedline{3}(120,120)(120,180)
 \drawline(60,0)(0,60)(120,180)
 \drawline(180,0)(240,60)(180,120)
 \thicklines
 \drawline(30,30)(120,120)
\end{picture}
\end{center}
\caption{Penrose diagram of a shell spacetime containing a piece of Minkowski
  spacetime with the second orientation. The star lines are the singularities,
  the dotted line is the regular center, the thick lines are the shells and
  the dashed lines are the Schwarzschild horizons.}\label{fig:oppos}
\end{figure}

The orientation of Minkowski space ($M_l = 0$) deserves a special attention.
There are two {\em physically different} space orientations of it: that with
the infinity right and the center left and the opposite one. A shell spacetime
constructed from a piece of Minkowski spacetime oriented in the latter way and
a piece of Schwarzschild spacetime is shown in Fig.~\ref{fig:oppos}. However,
the topology of Cauchy surfaces of the spacetime shown in Fig.~\ref{fig:oppos}
is ${\mathbf R} \times S^2$ instead of  ${\mathbf R}^3$ as required, and the
energy density of the shell there must be negative. 

This follows from a Lemma shown in \cite{DtH}:
\begin{lem}
Suppose that i) a null shell forms a common boundary of two Schwarz\-schild
spacetimes, ii) the spacetime with mass $M_2$ lies to the right and that with
$M_1$ to the left, and iii) the shell has positive energy density. Then either
the shell intersects ${\mathcal I}^+ \cup {\mathcal I}^-$ of the right scri
in both spacetimes and 
\[
  M_2 - M_1 > 0\ ,
\]
or the shell intersects ${\mathcal I}^+ \cup {\mathcal I}^-$ of the left scri
in both spacetimes and 
\[
  M_2 - M_1 < 0\ ,
\]
or the shell does not intersects ${\mathcal I}^+ \cup {\mathcal I}^-$ anywhere
and 
\[
  M_2 - M_1 = 0\ .
\]
\end{lem}
We assume, therefore, that the Minkowski spacetime is always oriented so that
the regular center of the constructed spacetime is left, and the infinity is
right.

In $\overline{\mathcal M}_K$, $K=m,r$, the so-called Double-Null
Eddington-Finkelstein (DNEF) coordinates are based on four functions $U_K^\pm$
and $V_K^\pm$. The function $U_K^\alpha$ is defined by
\begin{equation}
  U_K^\alpha := \alpha\left(T_K - R_K - 2M_K\ln\left|\frac{R_K}{2M_K} -
  1\right|\right)
\label{60-1}
\end{equation}
for $\alpha = \pm 1$. The domain of $U_K^+$ is the past of the
out-going Schwarzschild horizon $H_K^{\text{out}}$, that of $U_K^-$ the future
of $H_K^{\text{out}}$. Similarly,
\begin{equation}
  V_K^\beta := \beta\left(T_K + R_K +
  2M_K\ln\left|\frac{R_K}{2M_K} - 1\right|\right)\ , 
\label{60-3}
\end{equation}
the domain of $V_K^+$ being the future of the in-going Schwarzschild horizon
$H_K^{\text{in}}$, and that of $V_K^-$ the past of $H_K^{\text{in}}$.
Any pair of such functions define coordinates in that quadrant of
$\overline{\mathcal M}_K$ where their domains overlap. We denote these
quadrants by the pairs of signs $(\alpha\beta)$: $(++)$, $(+-)$, $(-+)$ and
$(--)$. 

The metric in ${\mathcal M}_K$ has the form
\begin{equation}
  ds^2 = -A_KdU_KdV_K + R_K^2d\Omega^2\ ,
\label{61-1}
\end{equation}
where
\begin{equation}
  A_K = \left|1-\frac{2M_K}{R_K}\right| = \alpha\beta
  \left(1-\frac{2M_K}{R_K}\right) 
\label{61-2}
\end{equation}
and, in each quadrant $(\alpha\beta)$,
\begin{equation}
  R_K = 2M_K\kappa\left[\alpha\beta\exp\left(\frac{-\alpha
  U_K^\alpha + \beta V_K^\beta}{4M_K}\right)\right]\ . 
\label{61-3} 
\end{equation}
Here, $\kappa$ is the well-known Kruskal function defined by its inverse:
\begin{equation}
  \kappa^{-1}(x) := (x - 1)e^x\ .
\label{star}
\end{equation}

Our motivation to work with these double-null coordinates is that, in terms of
them, many expressions and calculations, especially in the canonical part of
the formalism in papers II and III, significantly simplify. This may be due to
the simple relation of the DNEF coordinates to the time at infinity.

Observe that the formulae (\ref{61-1})--(\ref{61-3}) are all valid even if
$M_K = 0$ because\[
  \lim_{M \rightarrow 0} 2M\kappa\left[\exp\left(\frac{-U+V}{4M}\right)\right]
  = \frac{-U+V}{2}\ .
\]
There is then, of course, only one ``quadrant'', (++).

The notation settled, let us turn to the construction proper. The first step
is to choose the trajectory of the shell $S_1$ in $\overline{\mathcal
  M}_l$ which must be an in-going spherically symmetric null hypersurface. It
is defined by the equation 
\[
  V_l^+ = v_{l1}\ ,
\]
where $v_{l1} \in (-\infty,\infty)$ is a constant (the first parameter). The
second step is to choose the trajectories of $S_1$ and $S_2$ in
$\overline{\mathcal M}_m$. They must both be in-going null hypersurfaces, and
so have the equations
\[
  V_m^+ = v_{m1}
\]
and
\[
  V_m^+ = v_{m2}\ ,
\]
where $-\infty < v_{m1} < v_{m2} < \infty$ are two further parameters. The
third step is to choose the trajectory of $S_2$ in $\overline{\mathcal M}_r$:
\[
  V_r^+ = v_{r2}\ ,
\]
where $v_{r2} \in (-\infty,\infty)$ is another parameter.

The fourth step is to cut the pieces ${\mathcal M}_l$, ${\mathcal M}_m$ and
${\mathcal M}_r$ along the chosen boundaries from $\overline{\mathcal M}_l$,
$\overline{\mathcal M}_m$ and $\overline{\mathcal M}_r$ and paste them
together to make the solution spacetime $\mathcal M$. The pasting is defined
by the requirement that points with the same value of the Schwarzschild
coordinate $R$ coincide. This is always (i.e., for any allowed values of $M_m$,
$M_r$, $v_{l1}$, $v_{m1}$, $v_{m2}$ and $v_{r2}$) possible and unique, and it
closes the construction.

The parameters chosen on the way are $\eta_1$, $\eta_2$, $M_m$, $M_r$,
$v_{l1}$, $v_{m1}$, $v_{m2}$ and $v_{r2}$. Which of them describe physical
properties of the solution and which of them are gauge? It is clear that
$\eta_1$, $\eta_2$, $M_m$ and $M_r$ are physical, even geometric parameters.
The parameter $v_{m2} - v_{m1}$ determines, together with $M_m$, the geometry
between the two shells, so it is a geometric parameter, too.

The three parameters $M_m$, $M_r$ and $v_{m2} - v_{m1}$ determine the geometry
of the solution uniquely. Indeed, all pieces ${\mathcal M}_l(v_{l1})$ cut from
the Minkowski space $\overline{\mathcal M}_l$ along the curve $V_l^+ = v_{l1}$
for some value $v_{l1} \in (-\infty,\infty)$ are isometric to each other; all
pieces ${\mathcal M}_m(v_{m1},v_{m2})$ cut from $\overline{\mathcal M}_m$ with
a fixed mass $M_m$ and difference $v_{m2} - v_{m1}$ are isometric, and all
pieces ${\mathcal M}_r(v_{r2})$ are so too.  

However, the position of the system with respect to a frame at infinity is
also a physical parameter. In fact, $v_{r2}$ is the value of the advanced time
when the second shell is sent in; it is measurable by the observers near $i^0$
of $\mathcal M$ and can serve as the fourth observable.  In fact, the variable
$v_{r2} - v_{m2} + v_{m1}$ is also a frame-position observable and can be
chosen as one of our parameters instead of $v_{r2}$.

In this way, we have arrived at four coordinates describing the physical phase
space for case A ($\eta_1 = \eta_2 = -1$):
\[
  M_m, v_{m2}-v_{m1}, M_r, v_{r2}\ .
\]
Their domains are:
\begin{alignat*}{2}
  M_m & \in (0,\infty)\ , & \qquad M_r & \in (M_m,\infty)\ , \\
  v_{m2}-v_{m1} & \in (0,\infty)\ , & \qquad v_{r2} & \in (-\infty,\infty)\ .
\end{alignat*}
The remaining parameters $v_{l1}$ and $v_{m1}$ are then a kind of gauge.

The inequalities $M_m > 0$ and $M_r > M_m$ are a consequence of the
requirement that the energy density of the shells is positive
(cf.~\cite{DtH}). The manifold structure of this part of the physical phase
space, as well as the physical meaning of its points, are now both
well-defined.

Case B can be considered as time reversal of case A. The spacetime ${\mathcal
  M}_l$ is flat and has a regular center $c$. The first shell hypersurface is
$S_1$. The total energy of the shell is $M_m$, and it is simultaneously the
mass parameter of the Schwarzschild spacetime ${\mathcal M}_m$. The second
shell lies at $S_2$ and has total energy $M_r - M_m$. The Schwarzschild
spacetime ${\mathcal M}_r$ has mass parameter $M_r$.

The construction of the solution is entirely analogous to that of case A. The
shell trajectories are
\begin{alignat*}{2}
  U_l^+ & = u_{l1}\ , & \qquad U_m^+ & = u_{m1}\ , \\
  U_m^+ & = u_{m2}\ , & \qquad U_r^+ & = u_{r2}\ .
\end{alignat*}
The physical parameters are
\[
  M_m, u_{m2}-u_{m1}, M_r, u_{r2}\ ,
\]
and they have domains
\begin{alignat*}{2}
  M_m & \in (0,\infty)\ , & \qquad M_r & \in (M_m,\infty)\ , \\
  u_{m2} - u_{m1} & \in (0,\infty)\ , & \qquad u_{r2} & \in (-\infty,\infty)\ .
\end{alignat*}
The parameters $u_{l1}$ and $u_{m1}$ are again a kind of gauge.

\subsection{Crossing shells}
Cases C and C' can be tackled simultaneously if we work with out-going and
in-going shells instead of $s$-numerated ones. 

In this case, the shells have to intersect each other. This is a consequence
of the assumption that a piece of Minkowski spacetime contains a regular
center to the left. Hence, the shells separate each solution spacetime
$\mathcal M$ in four subspacetimes, cf.~Fig.~\ref{fig:C++}: The flat
subspacetime ${\mathcal M}_l$ (left), the subspacetime ${\mathcal M}_d$ 
(down) with Schwarzschild mass $M_d$, the subspacetime ${\mathcal M}_u$ 
(up) with mass $M_u$, and the subspacetime ${\mathcal M}_r$ (right) with 
mass $M_r$. The shells start with
total energies $M_r - M_d$ and $M_d$. After they cross at the point $O$ with
the value $r$ of Schwarzschild radius, they change their total energies to
$M_u$ and $M_r - M_u$. These parameters satisfy the relation
\begin{equation}
  r(r - 2M_r) = (r - 2M_u)(r - 2M_d)\ .
\label{16.2}
\end{equation}
Eq.~(\ref{16.2}) follows from the requirement of continuity at the crossing
point $O$ and has been derived in \cite{DtH} and \cite{Red}. The encounter
leads to exchange of energy between the shells, some energy passing from the
out-going to the in-going shell.

The construction of the solution in this case uses the same method as in the
case A but is more complicated. There are now four intershell spacetimes
${\mathcal M}_K$, $K = l,u,d,r$, four (properly oriented) extensions
$\overline{\mathcal M}_K$, three non-trivial masses, and the spacetimes
${\mathcal M}_K$ are wedges with vertices $O_K$ of radius $r$ rather than
stripes. The arrangement of the spacetimes ${\mathcal M}_K$ around the vertex
is illustrated by Fig.~\ref{fig:C++}.

In fact, the geometry of the solution is completely determined by the three
masses $M_u$, $M_d$ and $M_r$ because Eq.~(\ref{16.2}) implies that
\begin{equation}
  r = \frac{2M_uM_d}{M_u + M_d - M_r}\ .
\label{66-1}
\end{equation}

Important restrictions on the domains of the masses are
\begin{equation}
  M_u > 0\ ,\quad M_d > 0\ ,\quad M_r > 0\ .
\label{67-1}
\end{equation}
The first two are equivalent to the requirement that the energy density of the
shells is positive. The last one guarantees that the total energy of the
system is positive. This follows from the Positive Mass Theorem \cite{witten},
the topology of Cauchy surfaces being ${\mathbf R}^3$, and the positivity of
the energy density of the shells. Within this scenario, Eq.~(\ref{66-1})
implies
\begin{equation}
  M_r < M_u + M_d \ .
\label{67-2}
\end{equation}

Cases C, C' are different from cases A, A' and B, B' in that they contain
bound states of the external shell in the field of the internal one. This
manifests itself in the total mass of the external shell, $M_r - M_u$ or $M_r
- M_d$ being negative. There are four generic subcases depending on whether
both external shells are unbound, the out-going one is bound, the in-going is
bound, or both are bound. Between bound and unbound cases, there are five
boundary ones with one or both shells being marginally bound (lying at
Schwarzschild horizons).

\begin{figure}[t]
\begin{center}
\scalebox{.7}{
\begin{picture}(460,460)
 \put(424,40){$M_d$} \put(0,40){$M_u$} \put(65,410){$M_4$} 
 \put(0,370){$M_r=M$} \put(60,20){$M_d=0$} \put(380,20){$M_d=M$}  
 \put(380,0){$M_u=0$} \put(60,0){$M_u=M$} \put(213,310){C$_{++}$}
 \put(133,310){C$_{0+}$} \put(293,310){C$_{+0}$} \put(120,204){C$_{-+}$}
 \put(240,204){C$_{00}$} \put(320,204){C$_{+-}$} \put(133,110){C$_{-0}$}
 \put(213,110){C$_{--}$} \put(293,110){C$_{0-}$}
 \put(220,210){\circle{4}}
 \drawline(60,50)(218,208) \drawline(222,212)(380,370)(60,370)(218,212)
 \drawline(222,208)(380,50)(380,370)
 \put(60,50){\vector(1,0){360}} \put(60,50){\vector(-1,0){40}}
 \put(60,50){\vector(0,1){360}}
\end{picture}
}
\end{center}
\caption{Cut $M_u + M_d = M$ through the positive octant of the $M_u$, $M_d$
 and $M_r$ space}\label{fig:phsp}
\end{figure}
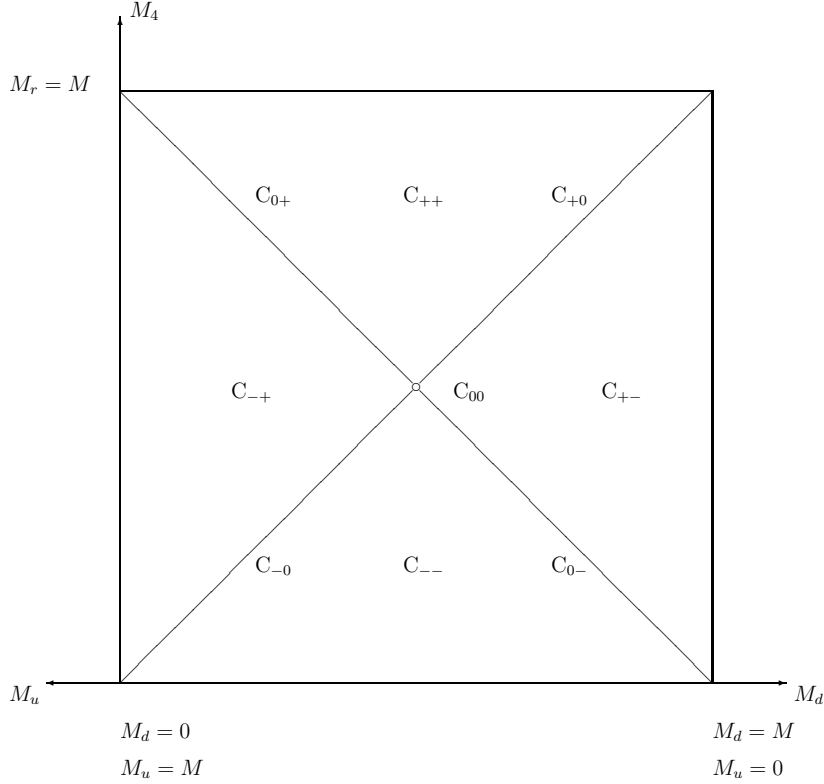

We shall denote the subcases of case C by C$_{ab}$, where
\begin{equation}
  a := \text{sgn}(M_r - M_u)\ ,\quad b := \text{sgn}(M_r - M_d)\ ,
\label{84-1}
\end{equation}
so that the values of $a$ and $b$ are $+$, $-$ or $0$. Case C' is geometrically
identical to case C, but the shells are swapped: the in-going shell becomes
the second one. The position of different subcases in the physical phase space
is illustrated by Fig.~\ref{fig:phsp}.

Given the masses $M_l = 0$, $M_u$, $M_d$ and $M_r$, the four Schwarzschild
spacetimes $\overline{\mathcal M}_l$, $\overline{\mathcal M}_u$,
$\overline{\mathcal M}_d$ and $\overline{\mathcal M}_r$ are determined. To
perform the construction, we need to know how the wedges ${\mathcal M}_l$, 
${\mathcal M}_u$, ${\mathcal M}_d$ and ${\mathcal M}_r$ are to be cut from
them. The orientation of the wedges is fixed by their relative position with
respect to the shells. Hence ${\mathcal M}_l$ opens left, ${\mathcal M}_d$
to the past, ${\mathcal M}_u$ to the future and ${\mathcal M}_r$ right. Then
each wedge is determined just by the coordinates of its vertex.

To find in which quadrant of $\overline{\mathcal M}_K$ the vertex $O_K$  lies
requires two steps. The first is based on Eq.~(\ref{66-1}). Two relations
between the vertex radius $r$ and differences of masses follow from it:
\begin{equation}
  r - 2M_r = \frac{2(M_r-M_u)(M_r-M_d)}{M_u + M_d - M_r}\ ,
\label{67-5}
\end{equation}
and 
\begin{equation}
  r - 2M_K = \frac{2M_K(M_r-M_K)}{M_u + M_d - M_r}\ 
\label{67-6}
\end{equation}
for $K = u,d$. Eq.~(\ref{67-6}) implies that $r > 2M_d$ if the in-going shell
is unbound, and that $r < 2M_d$ if it is bound. Similarly, $r > 2M_u$ ($r <
2M_u$) implies unboundedness (boundedness) for the out-going shell.
Eq.~(\ref{67-5}) requires that $r > 2M_r$ if none, or both, of the shells are
bound, and that $r < 2M_r$ if just one shell is bound. Thus, only two
quadrants from four remain to be eligible. To find the quadrant uniquely, we
use the so-called ``argument of matching divergences'' in the second step.

At each spherically symmetric null hypersurface in Schwarzschild spacetime,
the Schwarzschild radius either increases (divergence) or decreases
(convergence) in future direction or it is a part of one of the horizons. More
specifically, $U^+ =$ const and $V^- = $ const always diverge and $U^- =$
const and $V^+= $ const always converge for finite values of the constants. If
two null hypersurfaces are to be pasted together as in our construction, then
they must either both converge or both diverge or both have zero divergence
(lie on horizons).

The out-going shell in the past of the vertex $O_l$ must diverge in Minkowski
space $\overline{\mathcal M}_l$ and hence also in $\overline{\mathcal
  M}_d$, and it therefore lies in the domain of the function $U_d^+$. But this
domain contains only two quadrants, $(++)$ where $r > 2M_d$ and $(+-)$ where $r
< 2M_d$. Consequently, $O_d \in (++)$ ($O_d \in (+-)$) if the in-going shell
is unbound (bound). Similarly, $O_u \in (++)$ ($O_u \in (-+)$) if the
out-going shell is unbound (bound). In the marginally bound cases, the
vertices lie at the corresponding horizons.

\begin{figure}
\begin{center}
\begin{picture}(240,160)
 \dashline[50]{5}(60,60)(120,0)
 \dashline[50]{5}(100,60)(200,160) 
 \dottedline[$\star$]{9}(0,160)(200,160)
 \dottedline[$\star$]{9}(0,0)(120,0)
 \dottedline{3}(0,0)(0,160)
 \drawline(120,0)(240,120)(200,160)
\thicklines
 \drawline(0,0)(160,160) \drawline(0,160)(140,20)
\end{picture}
\end{center}
\caption{Penrose diagram of subcase C$_{-+}$. The dotted line is the regular
 center, the star lines are the singularities, the dashed lines are
 Schwarzschild horizons and the thick lines are the shells.  Subcase C$_{+-}$
 can be considered as the time reversal of C$_{-+}$.} \label{fig:C-+}
\end{figure}
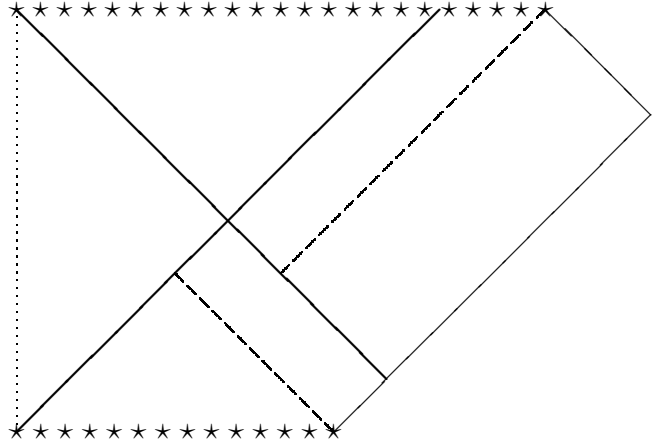

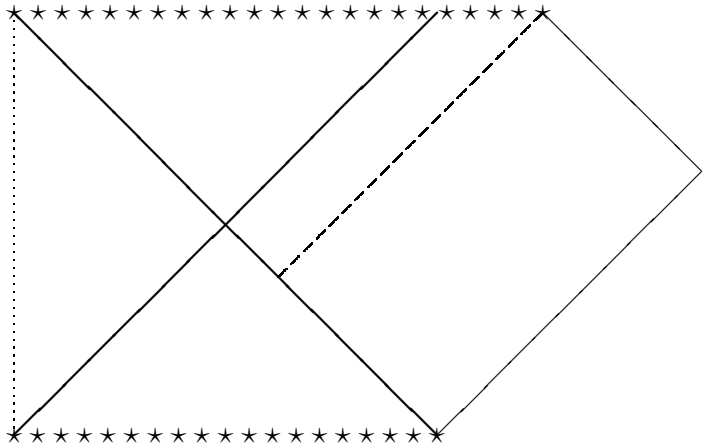
\begin{figure}
\begin{center}
\begin{picture}(260,160)
 \dashline[50]{5}(100,60)(200,160)
 \dottedline[$\star$]{9}(0,160)(200,160)
 \dottedline[$\star$]{9}(0,0)(160,0)
 \dottedline{3}(0,0)(0,160)
 \drawline(160,0)(260,100)(200,160)
\thicklines
 \drawline(0,0)(160,160) \drawline(0,160)(160,0)
\end{picture}
\end{center}
\caption{Penrose diagram of subcase C$_{-0}$. The dotted line is the regular
 center, the star lines are the singularities, the dashed line is
 Schwarzschild horizon and the thick lines are the shells.  Subcase C$_{0-}$
 can be considered as the time reversal of C$_{-0}$.} \label{fig:C-0}
\end{figure}

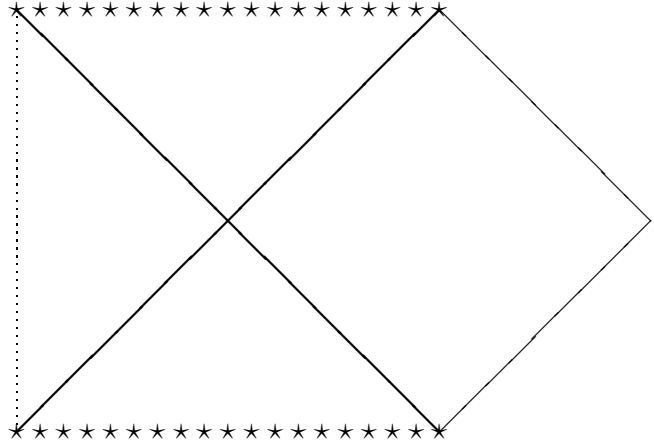
\begin{figure}
\begin{center}
\begin{picture}(240,160)
 \dottedline[$\star$]{9}(0,160)(160,160)
 \dottedline[$\star$]{9}(0,0)(160,0)
 \dottedline{3}(0,0)(0,160)
 \drawline(160,0)(240,80)(160,160)
\thicklines
 \drawline(0,0)(160,160) \drawline(0,160)(160,0)
\end{picture}
\end{center}
\caption{Penrose diagram of subcase C$_{00}$. The dotted line is the regular
 center, the star lines are the singularities, and the thick lines are the
 shells.} \label{fig:C00}
\end{figure}

\begin{figure}
\begin{center}
\begin{picture}(280,160)
 \dashline[50]{5}(100,60)(200,160)
 \dashline[50]{5}(100,100)(200,0) 
 \dottedline[$\star$]{9}(0,160)(200,160)
 \dottedline[$\star$]{9}(0,0)(200,0)
 \dottedline{3}(0,0)(0,160)
 \drawline(200,0)(280,80)(200,160)
\thicklines
 \drawline(0,0)(160,160) \drawline(0,160)(160,0)
\end{picture}
\end{center}
\caption{Penrose diagram of subcase C$_{--}$. The dotted line is the regular
 center, the star lines are the singularities, the dashed lines are
 Schwarzschild horizons and the thick lines are the shells.} \label{fig:C--}
\end{figure}
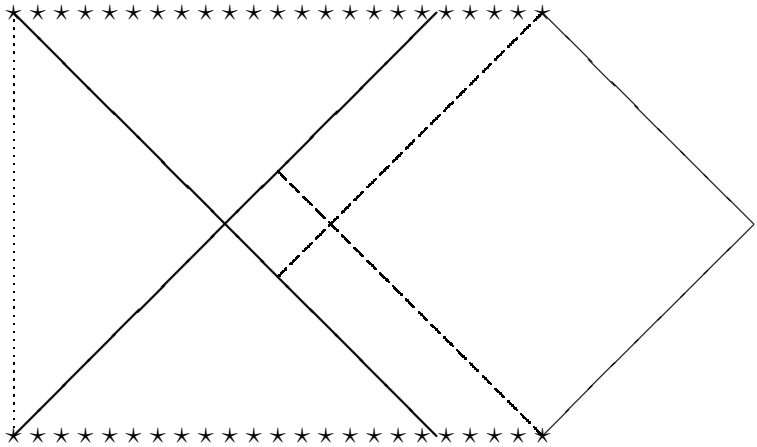

Applying the argument of matching divergences to the wedge ${\mathcal M}_r$,
we find that the in-going shell converges (diverges) if it is unbound (bound),
and the out-going shell diverges (converges) if it is unbound (bound). This
leads to $O_r$ lying in the quadrant $(++)$ if both shells are unbound, in
$(-+)$ if the out-going shell is bound and the in-going one unbound, in $(+-)$
if the out-going shell is bound and the in-going one unbound, and in $(--)$
if both shells are bound. The resulting spacetimes are shown by
Figs.~\ref{fig:C++}--\ref{fig:C--}. The marginally bound cases are also
included. 

\begin{figure}
\begin{center}
\begin{picture}(200,160)
 \put(71,96){${\mathcal M}_u$} \put(163,126){${\mathcal I}^+$}
 \put(165,29){${\mathcal I}^-$} \put(50,74){${\mathcal M}_l$}  
 \put(95,75){${\mathcal M}_r$} \put(73,54){${\mathcal M}_d$}
 \dashline[50]{5}(60,100)(120,160)
 \dashline[50]{5}(60,60)(120,0) 
 \dottedline[$\star$]{9}(0,160)(120,160)
 \dottedline[$\star$]{9}(0,0)(120,0)
 \dottedline{3}(0,0)(0,160)
 \drawline(120,0)(200,80)(120,160)
\thicklines
 \drawline(0,0)(140,140) \drawline(0,160)(140,20)
\end{picture}
\end{center}
\caption{Penrose diagram of subcase C$_{++}$. The dotted line is the regular
 center, the star lines are the singularities, the dashed lines are
 Schwarzschild horizons and the thick lines are the shells.} \label{fig:C++}
\end{figure}
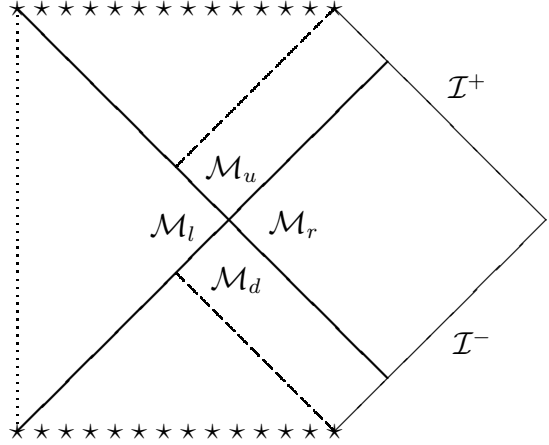

\begin{figure}
\begin{center}
\begin{picture}(220,160)
 \dashline[50]{5}(60,60)(120,0)
 \dottedline[$\star$]{9}(0,160)(160,160)
 \dottedline[$\star$]{9}(0,0)(120,0)
 \dottedline{3}(0,0)(0,160)
 \drawline(120,0)(220,100)(160,160)
\thicklines
 \drawline(0,0)(160,160) \drawline(0,160)(140,20)
\end{picture}
\end{center}
\caption{Penrose diagram of subcase C$_{0+}$. The dotted line is the regular
 center, the star lines are the singularities, the dashed line is
 Schwarzschild horizon and the thick lines are the shells. Subcase C$_{+0}$
 can be considered as the time reversal of C$_{0+}$.} \label{fig:C0+}
\end{figure}
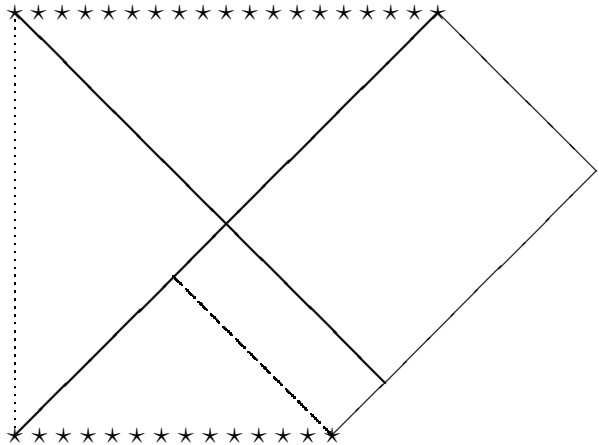

The construction is determined by specifying the coordinates of the four
vertices $O_K$ in the spacetimes $\overline{\mathcal M}_K$. The parameters
$u_K$ and $v_K$, $K=l,d,u,r$, are defined as follows. In all cases, $O_l$ has
finite coordinates and $u_l = U_l^+$ and $v_l = V_l^+$. For the vertex $O_d$,
we always have $u_d = U_d^+$. On the other hand, if the in-going shell is
unbound $v_d = V_d^+$, if it is bound $v_d = V_d^-$, and if it is marginally
bound $V_d^+ = - V_d^- = \infty$. Similarly for other vertices; we can
summarize the definition of the parameters as follows. For case C$_{ab}$ with
$a \neq 0$ and $b \neq 0$, it holds that
\begin{alignat*}{2}
  u_d & = U_d^+\ , & \qquad v_d & = V_d^b\ ,  \\
  u_u & = U_u^a\ , & \qquad v_u & = V_u^+\ ,  \\
  u_r & = U_r^a\ , & \qquad v_r & = V_r^b\ .
\end{alignat*}
For the marginally bound cases, if $a = 0$ then $U_u^\pm$ and $U_r^\pm$
diverge, and if $b = 0$ then $V_d^\pm$ and $V_r^\pm$ diverge. The parameters
$u_d$ and $v_u$ are always defined by $u_d = U_d^+$ and $v_u = V_u^+$.

Eq.~(\ref{61-3}) shows that some combinations of the two vertex coordinates is
determined by the masses and cannot serve as independent parameters. In case
C$_{ab}$, if $a \neq 0$ and $b \neq 0$, we have
\begin{eqnarray}
  v_l - u_l & = & 2r\ , \nn \\
  bv_d - u_d & = & 4M_d\ln\left[b\kappa^{-1}\left(\frac{r}{2M_d}\right)\right]\
  , \label{13/1} \\  
  v_u - au_u & = & 4M_u\ln\left[a\kappa^{-1}\left(\frac{r}{2M_u}\right)\right]\
  , \label{13/2} \\  
  bv_r - au_r & = &
  4M_r\ln\left[ab\kappa^{-1}\left(\frac{r}{2M_r}\right)\right]\ 
  , \label{13/3}
\end{eqnarray}
where $r$ is given by Eq.~(\ref{66-1}). Hence, only one parameter of each pair
$(u_K, v_K)$ is free (if it is finite), and then it runs through the whole
range $(-\infty, \infty)$ within each case. 

Let us turn to the question which parameter is to be chosen for the purpose of
providing information about the position of the shell pair with respect to
infinity.  The most straightforward choice is either $u_r$ or $v_r$. Clearly,
$u_r$ and $v_r$ are even directly observable from infinity as retarded or
advanced time coordinates of the out-going or in-going shell, as long as the
shells are not bound or marginally bound.

It seems, however, that a conflict between two different notions of
observability emerges here. On the one hand, there is observability in the
canonical sense where the whole Cauchy surface is observable, and on the other
hand there is observability from the asymptotic infinity. This does not
concern the non-geometric observables only. Even some geometric observables
seem not to be observable or manipulable from the infinity in the bound cases.
For example, $bv_d - u_d$ $(b=-)$ is not observable in cases C$_{0-}$ C$_{+-}$
C$_{--}$. This is an interesting point that must have some relevance for our
program of reducing the collapse to a scattering problem. It cannot, however,
be dealt with in the present paper; only the observability in the canonical
sense is considered here.

In the generic cases $a \neq 0$ and $b \neq 0$, both $u_r$ and $v_d$ are
regular and any of them can be chosen. They are related by
\begin{equation}
  v_r = abu_r + 4aM_r\ln\left[ab\kappa^{-1}\left(\frac{r}{2M_r}\right)\right]\
  , 
\label{6/0}
\end{equation}
where $r$ is defined by Eq.~(\ref{66-1}). In addition, $u_r$ is well-defined
in the cases C$_{+0}$ and C$_{-0}$, $v_r$ in C$_{0+}$ and C$_{0-}$. If we
exclude the case C$_{00}$, then the rest of the physical phase space can be
covered by two coordinate charts. The first chart is defined by the
functions $M_d$, $M_u$, $M_r$ and $u_r$; it covers the cases C$_{++}$,
C$_{+0}$, C$_{+-}$, C$_{-+}$, C$_{-0}$ and C$_{--}$. The second is defined by
$M_d$, $M_u$, $M_r$ and $v_r$ and covers the cases C$_{++}$, C$_{0+}$,
C$_{+-}$, C$_{-+}$, C$_{0-}$ and C$_{--}$. In the overlapping regions C$_{++}$,
C$_{+-}$, C$_{-+}$ and C$_{--}$, there is a unique transformation. In cases
C$_{0+}$, C$_{+0}$, C$_{0-}$ and C$_{-0}$ we have only one choice because one
of the two parameters diverges. 

In case C$_{00}$, both parameters $u_r$ and $v_r$ diverge. From the point of
view of infinity, all solutions of the case are static; their position with
respect to a frame at infinity is not defined: no property measurable at
infinity changes if the frame at infinity is time shifted. However, complete
solutions of case C$_{00}$ do not possess any additional symmetry in
comparison to other cases because ${\mathcal M}_l$, ${\mathcal M}_u$ and
${\mathcal M}_d$ do not. The constraint surface of the ADM formalism is
regular (cf.~\cite{moncrief}), and there is a well defined pull back of the
symplectic form to the constraint surface there. 

A regular coordinate system in the phase space is constructed in paper III.
It cannot be constructed in the present paper because it has to fulfill more
conditions than just boundedness of the coordinate functions at all points of
the space. The most important is of course the regularity of the symplectic
form. We certainly could replace $v_r$ by functions of $v_r$, $M_d$, $M_u$ and
$M_r$ that would be regular at C$_{00}$, but there is no point in doing this
until we know the symplectic form.

\section{Choice of gauge}
\label{sec:gauge}
By a choice of gauge, a unique background manifold $\mathsf{M}$ is specified
and each solution can be described as a set of fields and branes on
$\mathsf{M}$, dependent on both the physical phase space coordinates and those
on $\mathsf{M}$ (cf.~\cite{H-Kij} and \cite{H-Kie}). The fields have to satisfy
certain conditions. Thus, for the case of gravitating shells, the metric is
required to be continuous, piecewise smooth and to satisfy some boundary
conditions at the regular center and at the infinity (\cite{H-Kie}). A
continuous metric determines uniquely a class of $C^1$ coordinate systems. Any
gauge is required to be $C^1$. In this sense, no gauge has been chosen as yet.

A point important for the subsequent calculations is that there is a gauge
choice for each case and that there is a well defined transformation between
such gauge and the description of solutions of the case as given in the
preceding section. This transformation is singular at some points (some
transformation functions are divergent and some have step discontinuities),
and it depends on more parameters that the dimension of the physical phase
space has. However, the transformation is, locally, very much like a gauge
transformation. Let us show it for case A.

Our construction of a regular gauge is based on the following interesting
property of the DNEF coordinates. 
\begin{lem} Let $p$ be a point of a shell, let $(X_1,Y_1)$ be some DNEF
  coordinate pair in a neighborhood $\mathcal U$ of $p$ left from the shell
  and $(X_2,Y_2)$ that right from the shell. Let the shell points in $\mathcal
  U$ satisfy the equations $X_1 = X_1(p)$ and $X_2 = X_2(p)$. Let the
  Schwarzschild radial coordinate $R = R_n(X_n,Y_n)$ satisfy $R_n(X_n,Y_n)
  \neq 2M_n$ in $\mathcal U$ for both $n = 1,2$. Let, finally, the functions
  $X$ and $Y$ be defined right from the shell by
\begin{equation}
  X = X_2 - X_2(p) +X_1(p)
\label{94-1}
\end{equation}
and 
\begin{equation}
  R_2(X_2(p),Y_2(Y)) = R_1(X_1(p), Y)\ ,
\label{94-2}
\end{equation}
and left from the shell by
\begin{equation}
  X = X_1\ , \quad Y = Y_1\ .
\label{94-3}
\end{equation}
Then the functions $X$ and $Y$ form a $C^1$ coordinate system in $\mathcal U$.
\end{lem} 
{\bf Proof} An observation that is crucial for the proof is the following:
$X_1$ and $X_2$ must always be Eddington-Finkelstein functions of the same
type. This follows immediately from the argument of matching divergences.
Hence, there are four cases. These cases define the sign $\zeta$ by the
following table:
\begin{center}
\begin{tabular}{|l|l|l|}\hline
$X_1$ & $X_2$ & $\zeta$ \\ \hline
$U_1^+$ & $U_2^+$ & $+$ \\
$U_1^-$ & $U_2^-$ & $-$ \\
$V_1^+$ & $V_2^+$ & $-$ \\
$V_1^-$ & $V_2^-$ & $+$ \\ \hline
\end{tabular}\end{center}
Let us first consider the usual Eddington-Finkelstein coordinates $X_n$ and
$R$ to both sides of the shell; the metric is
\[
  ds^2 = -\left(1 - \frac{2M_n}{R}\right)dX_n^2 - 2\zeta dX_ndR +
  R^2d\Omega^2\ .
\]
The transformation to $(X_n,Y_n)$ can be performed by setting $R =
R_n(X_n,Y_n)$ and we obtain in this way
\[
  ds^2 = -\left[\left(1-\frac{2M_n}{R}\right) + 2\zeta\frac{\partial
  R_n}{\partial X_n}\right] dX_n^2
  -2\zeta\frac{\partial R_n}{\partial Y_n}dX_ndY_n + R^2_nd\Omega^2\ .
\]
It follows then that
\[
  \frac{\partial R_n}{\partial X_n} =
  -\frac{\zeta}{2}\left(1-\frac{2M_n}{R}\right) 
\]
and
\begin{equation}
  \frac{\partial R_n}{\partial Y_n} = \frac{\zeta}{2}A_n\ ,
\label{96-1}
\end{equation}
where $A_n$ appears in the $(X_n,Y_n)$ metric:
\[
  ds^2 = - A_ndX_ndY_n + R_n^2d\Omega^2\ .
\]
Eq.~(\ref{94-2}) implies
\[
  \left(\frac{\partial R_1}{\partial Y_1}\right)_{\text{shell},Y_1=Y} =
  \left(\frac{\partial R_2}{\partial Y_2}\right)_{\text{shell}} \frac{\partial
  Y_2}{\partial Y} 
\]
or 
\[
  \frac{\partial Y_2}{\partial Y} = \frac{\left(\frac{\partial R_1}{\partial
  Y_1}\right)_{\text{shell},Y_1=Y}} {\left(\frac{\partial R_2}{\partial
  Y_2}\right)_{\text{shell}}}\ .
\]
Then, using Eq.~(\ref{96-1}), we have
\[
 \frac{\partial Y_2}{\partial Y} = \frac{A_1\vert_{\text{shell}}}
 {A_2\vert_{\text{shell}}}\ .
\]

The transformation (\ref{94-3}) gives, left from the shell,
\[
  ds^2 = -A_1(X,Y)dXdY + R_1^2(X,Y)d\Omega^2\ .
\]
Similarly, the transformation (\ref{94-1}) and (\ref{94-2}) leads to, right
from the shell,
\[
  ds^2 = -A_2(X,Y)\frac{A_1}{A_2}\vert_{\text{shell}}dXdY +
  R_2^2(X,Y)d\Omega^2\ . 
\] 
Thus, the metric is continuous at the shell, because $R$ is, QED.

Our gauge for case A can be described as follows:
\paragraph{Construction of the $V$ coordinate in ${\mathcal M}_l$, ${\mathcal
    M}_m$, ${\mathcal M}_r$:} Let 
\[
V := V_l^+ + v_{r2} - v_{m2} + v_{m1} - v_{l1}
\]
in ${\mathcal M}_l$,
\[
   V := V_m^+ + v_{r2} - v_{m2} 
\]
in ${\mathcal M}_m$ and 
\[
   V := V_r^+
\]
in ${\mathcal M}_r$. Then $V$ is a continuous function in $\mathcal M$. The
inverse transformation is
\begin{equation}
  V_l^+ := V - v_{r2} + v_{m2} - v_{m1} + v_{l1}
\label{88-1}
\end{equation}
for $V \in (-\infty,v_{r2} - v_{m2} + v_{m1})$,
\begin{equation}
  V_m^+ := V - v_{r2} + v_{m2}
\label{88-2}
\end{equation}
for $V \in (v_{r2} - v_{m2} + v_{m1},v_{r2})$ and 
\begin{equation}
  V_r^+ := V
\label{88-3}
\end{equation}
for $V \in (v_{r2},\infty)$.

The construction of the $U$ coordinate is more laborious:
\paragraph{Construction of the $U$ coordinate in ${\mathcal M}_l$:}
 In the left region, we set
\begin{equation}
  U := U_l^+  + v_{r2} - v_{m2} + v_{m1} - v_{l1}\ ,
\label{89-3}
\end{equation}
and the metric is given by 
\begin{equation}
  A_l(U,V) = 1\ ,
\label{89-4}
\end{equation} 
and
\begin{equation}
  R_l(U,V) = \frac{-U+V}{2}\ .
\label{89-5}
\end{equation}

\paragraph{Construction of the $U$ coordinate in ${\mathcal M}_m$:}
Across the first shell, we must have
\[
  R_l(U,V_l^+=v_{l1}) = R_m(U_m^+,V_m^+=v_{m1})
\]
for $U_m^+ \in (-\infty,\infty)$.  Using Eq.~(\ref{61-3}), we obtain the
relation between $U$ and $U_m^+$ along this part of the shell:
\begin{equation}
  U = v_{r2} - v_{m2} + v_{m1} - 4M_m\kappa\left[\exp\left(\frac{-U_m^+ +
  v_{m1}}{4M_m}\right)\right]\ .
\label{89-1}
\end{equation}
Along the part $U_m^- \in (-\infty,-v_{m1})$ of the first shell, we obtain in
an analogous way:
\begin{equation}
  U = v_{r2} - v_{m2} + v_{m1} - 4M_m\kappa\left[-\exp\left(\frac{U_m^- +
  v_{m1}}{4M_m}\right)\right]\ .
\label{89-2}
\end{equation}
Let us use the right hand sides of Eqs.~(\ref{89-1}) and (\ref{89-2}) as the
definitions of the coordinate $U$ everywhere in ${\mathcal M}_m$. Then $U$ is
continuous across the first shell. The inverse transformation is
\begin{equation}
  U_m^+(U) = v_{m1}
  -4M_m\ln\left[\kappa^{-1}\left(\frac{-U+v_{r2} - v_{m2} +
  v_{m1}}{4M_m}\right)\right]
\label{90-1}
\end{equation}
for $U \in (-\infty,U(H_m))$, where $H_m$ is the Schwarzschild horizon in
${\mathcal M}_m$, and 
\begin{equation}
  U_m^-(U) = v_{m1}
  -4M_m\ln\left[-\kappa^{-1}\left(\frac{-U+v_{r2} - v_{m2} + v_{m1}
  }{4M_m}\right)\right]
\label{90-2}
\end{equation}
for $U \in (U(H_m),v_{r2} - v_{m2} + v_{m1})$. $U(H_m)$ can easily be
calculated from Eq.~(\ref{89-1}) by setting $U_m^+ = +\infty$ yielding the
result $U(H_m) = v_{r2} - v_{m2} + v_{m1} - 4M_m$.  If we substitute either
Eqs.~(\ref{88-2}) and (\ref{90-1}) or Eqs.~(\ref{88-2}) and (\ref{90-2}) into
Eq.~(\ref{61-3}), we obtain in both cases that
\begin{equation}
  R_m =
  2M_m\kappa\left[\exp\left(\frac{V-v_{r2} + v_{m2} - v_{m1}}{4M_m}\right)
  \kappa^{-1}\left(\frac{-U+v_{r2}- v_{m2} + v_{m1}}{4M_m}\right)\right]\ .  
\label{90-3}
\end{equation}

The metric in ${\mathcal M}_m$ can be written in the form
\begin{equation}
  ds^2 = -A_m(U,V)dUdV + R_m^2(U,V)d\Omega^2;
\label{7.2}
\end{equation}
it is obtained from the Eddington-Finkelstein line element
\[
  ds^2 = -\left(1-\frac{2M_m}{R}\right)dV^2 + 2dVdR + R^2d\Omega^2
\]
by the transformation $R = R_m(U,V)$. This implies that
\begin{equation}
  A_m = -2\frac{\partial R_m}{\partial U}
\label{7.3}
\end{equation}
and 
\begin{equation}
  \frac{\partial R_m}{\partial V} = \frac{1}{2}\left(1 -
  \frac{2M_m}{R_m}\right). 
\label{7.4}
\end{equation}
Eqs.~(\ref{90-3})--(\ref{7.3}) determine the metric in ${\mathcal M}_m$ as
function of the coordinates $U$ and $V$ and parameters $M_m$, $v_{r2}$ and
$v_{m2}-v_{m1}$.

\paragraph{Construction of the $U$ coordinate in ${\mathcal M}_r$:}
The coordinate $U$ is determined in ${\mathcal M}_r$ from the condition of
continuity across the second shell,
\[
  R_m(U,v_{r2}) = R_r(U_r^+,v_{r2})
\]
for $U_r^+ \in (-\infty,\infty)$, and 
\[
  R_m(U,v_{r2}) = R_r(U_r^-,v_{r2})
\]
for $U_r^- \in (-\infty,-v_{r2})$, where $R_r$ is the Schwarzschild radial
coordinate in ${\mathcal M}_r$. A procedure analogous to that for ${\mathcal
  M}_m$ results in the transformation formulae:
\begin{multline}
  U_r^+(U) = v_{r2} - \\
  4M_r\ln\left(\kappa^{-1}\left\{\frac{M_m}{M_r}
  \kappa\left[\exp\left(\frac{v_{m2}-v_{m1}}{4M_m}\right)
  \kappa^{-1}\left(\frac{-U +
  v_{r2}-v_{m2}+v_{m1}}{4M_m}\right)\right]\right\}\right) 
\label{91-1}
\end{multline}
for $U \in (-\infty,U(H_r))$, and 
\begin{multline}
  U_r^-(U) = -v_{r2} + \\
  4M_r\ln\left(-\kappa^{-1}\left\{\frac{M_m}{M_r}
  \kappa\left[\exp\left(\frac{v_{m2}-v_{m1}}{4M_m}\right)
  \kappa^{-1}\left(\frac{-U +
  v_{r2}-v_{m2}+v_{m1}}{4M_m}\right)\right]\right\}\right)   
\label{91-2}
\end{multline}
for $U \in (U(H_r), U_{0r})$, where $H_r$ is the Schwarzschild horizon in
${\mathcal M}_r$, and $U_{0r}$ is the value of the coordinate $U$ for which
the second shell has zero radius,
\[
  U(H_r) = v_{r2} - v_{m2} + v_{m1} -
  4M_m\kappa\left[\exp\left(\frac{-v_{m2}+v_{m1}}{4M_m}\right)
  \kappa^{-1}\left(\frac{M_r}{M_m}\right)\right]\ ,
\]
and
\[
  U_{0r} = v_{r2} - v_{m2} + v_{m1} -
  4M_m\kappa\left[-\exp\left(\frac{-v_{m2}+v_{m1}}{4M_m}\right)\right]\ .
\]

Eqs.~(\ref{61-3}), (\ref{88-3}), (\ref{91-1}) and (\ref{91-2}) imply the
following line element:
\begin{equation}
  ds^2 = -A_r(U,V)dUdV + R^2_r(U,V)d\Omega^2,
\label{8.1}
\end{equation}
where
\begin{eqnarray}
  \lefteqn{R_r(U,V) =
  2M_r\kappa\left(\exp\left\{\frac{V-v_{r2}}{4M_r}\right\}\right.\times} \nn
  \\   
  && \left.\kappa^{-1}
  \left\{\frac{M_m}{M_r}\kappa\left[\exp\left(\frac{v_{m2}-v_{m1}}{4M_m}\right)
  \kappa^{-1}\left(\frac{-U +
  v_{r2}-v_{m2}+v_{m1}}{4M_m}\right)\right]\right\}\right)
\label{92-1}
\end{eqnarray}
for $U \in (-\infty,U_{0r})$, 
\begin{equation}
  A_r(U,V) = -2\frac{\partial R_r}{\partial U}\ ;
\label{92-2}
\end{equation}
and the identity 
\[
  \frac{\partial R_r}{\partial V} = \frac{1}{2}\left(1 -
  \frac{2M_r}{R_r}\right) 
\]
is also valid.
One easily verifies that the function $A(U,V)$ defined by $A(U,V) = A_i(U,V)$
in ${\mathcal M}_i$ is continuous across the shells. The function $R(U,V)$
defined by $R(U,V) = R_i(U,V)$ in ${\mathcal M}_i$ is continuous by
construction. The background manifold $\mathsf{M}$ can be defined by the
domains of coordinates $U$ and $V$:
\[
  U \in (-\infty,\infty)\ ,\quad V \in (U,\infty)\ ;
\]
the regular center part of the boundary is $V = U$, $U = -\infty$ is
${\mathcal I}^-$ and $V = \infty$ is ${\mathcal I}^-$. The solutions are
described as metric fields and shell trajectories on $\mathsf{M}$ that depend
on the parameters $M_m$, $M_r$, $v_{r2}$ and $v_{m2} - v_{m1}$.

The transformation between the coordinates $U$ and $V$ on one hand and
$U_l^+$, $V_l^+$, $U_m^\pm$, $V_m^+$, $U_r^\pm$ and $V_r^+$ on the other is
given by Eqs.~(\ref{88-1})--(\ref{88-3}), (\ref{89-3}), (\ref{90-1}),
(\ref{90-2}), (\ref{91-1}) and (\ref{91-2}). They depend on additional
parameters $v_{m1}$ and $v_{l1}$.

We observe that the transformation functions have step discontinuity at $V =
v_{r2} - v_{m2} + v_{m1}$ and $V = v_{r2}$, they are divergent at $U = U(H_m)$
for $V \in (v_{r2} - v_{m2} + v_{m1},v_{r2})$ and at $U = U(H_r)$ for $V \in
(v_{r2},\infty)$. In all other points, they represent a regular,
parameter-dependent gauge transformation. We shall, therefore, call
coordinates similar to $U_l^+$, $V_l^+$, $U_m^\pm$, $V_m^+$, $U_r^\pm$ and
$V_r^+$ {\em singular gauge}. One might be able to use singular gauges for
calculation of some gauge-invariant quantities. In fact, we shall do so in the
subsequent papers, II and III.

\section{Symmetries}
\label{sec:symm}
In Sec.~\ref{sec:solutions}, a complete account of the space of solutions has
been provided. In the present section, we are going to describe some relations
between different solutions in this space. First, we observe that some
solutions are isometric to each other. The isometries are of two types; time
shift and time reversal. Second, there are also conformally related solutions.
The time shifts and dilatations generate one-dimensional subfamilies of
solutions; the time reversal acts only inside pairs of solutions. The
symmetries are interesting for us mainly because they will be used for a
construction of the most interesting observables (such as, e.g., the
Hamiltonian) in the quantum theory (cf.~\cite{H})

The fact that physically different solutions are isometric may need some
comment: how can two isometric solutions be physically different? The cause
of such a difference is the asymptotically flat region. We consider just one
family of asymptotic observers common for the whole space of solutions. For
these observers, two shells such that one is sent earlier than the other
represent two physically different situations even if the shells are
completely isomorphic in all other properties. In fact, the time shifts and
the time reversal is the only remnant of the Bondi-Metzner-Sachs (BMS) group
$G_{\text{BMS}}$ that is non-trivial and preserves the spherical symmetry of
our model. In a general situation with asymptotically flat region, the whole
BMS group of transformations between standard asymptotic frames acts as a
group of symmetry. Indeed, the full diffeomorphism group $G$ (each element of
which sends any solution isometrically to another one, see \cite{H-Kie}) can
be split into the gauge and the symmetry groups. A subgroup $G_0$ of $G$ that
leaves, roughly speaking, all points at the infinity invariant ($G_0$ has to be
selected by a suitable fall-off condition) is the gauge group; the
factor group $G/G_0 = G_{\text{BMS}}$ is the group of symmetry.

Each symmetry $\Phi$ has well-defined actions in different spaces of interest.
First, $\Phi$ acts in the space of parameters that describe the solutions. The
action just tells us which pairs of solutions are related by $\Phi$. Next, if a
gauge is chosen, $\Phi$ also acts on the corresponding background manifold.
This action tells us which points of the two $\Phi$-related solution
spacetimes are mapped on each other by the corresponding diffeomorphism.
Third, $\Phi$ acts in the asymptotic space that is common to all solutions and
that represents the view of the asymptotic observers. Fourth, $\Phi$ also acts
on the phase space of the system. In fact, the first action is nothing but the
action on the physical (reduced) phase space; one only has to equip the space
of parameters with a symplectic structure. But there will also be an action on
the constraint surface, the points of which are initial data for solutions, or
even on the extended phase space. In this section, we shall study only the
first three actions.

\subsection{Time shift}
\label{sec:timesh}
Time shifts form a one-dimensional group of transformations. Let the group
parameter be denoted by $t$. Time shifts preserve the in- and out-going
character of the shells, so we can work case by case.

Consider case A. The coordinates on the physical phase space are $M_m$,
$v_{m2} - v_{m1}$, $M_3$ and $v_{r2}$, The three variables $M_m$,
$v_{m2} - v_{m1}$ and $M_r$ determine the geometry uniquely. Thus, solutions
that do not differ in the values of these parameters are isometric. Let us
define the action $\Phi_t^P$ of the $t$-shift isometry $\Phi_t$ on the
physical phase space by
\[
  \Phi^P_t(M_m,v_{m2} - v_{m1},M_r,v_{r2}) = (M_m,v_{m2} -
  v_{m1},M_r,v_{r2}+t)\ .
\]
This definition is suggested by the nature of the parameter $v_{r2}$. As the
advanced time of the external shell, it is measurable by the asymptotic
observers and it is shifted by $t$ if the asymptotic observers shift their
clocks by $-t$. 

In Sec.~\ref{sec:gauge}, a particular gauge is chosen for case A. A background
manifold ${\mathsf M}$ covered by coordinates $U$ and $V$ is well defined,
and each solution appears as a particular metric field
\begin{equation}
  ds^2 = -A(U,V;M_m,v_{m2} - v_{m1},M_r)dUdV + R^2(U,V;M_m,v_{m2} -
  v_{m1},M_r)d\Omega^2 
\label{99-1}
\end{equation}
on ${\mathsf M}$, where the functions $A$ and $R$ are defined in
Sec.~\ref{sec:gauge} by Eqs.~(\ref{89-4}), (\ref{89-5}), (\ref{90-3}),
(\ref{7.3}), (\ref{92-1}) and (\ref{92-2}). The trajectories of the shells in
${\mathsf M}$, 
\begin{equation}
  V = v_{r2} - v_{m2} - v_{m1}
\label{99-2}
\end{equation}
and
\begin{equation}
 V = v_{r2} 
\label{99-3}
\end{equation}
also depend on the parameters. Hence, there must be a unique map
$\Phi^{\mathsf M}_t : {\mathsf M} \mapsto {\mathsf M}$ for each $t$ such that
the metric and the shell trajectories given by the values of the parameters
$M_m$, $v_{m2} - v_{m1}$, $M_r$ and $v_{r2}$ are mapped into the metric and
trajectories for $M_m$, $v_{m2} - v_{m1}$, $M_r$ and $v_{r2} + t$. This is
done by the map
\begin{equation}
  \Phi^{\mathsf M}_t(U,V) = (U+t,V+t)\ .
\label{99-4}
\end{equation}
Indeed, Eqs.~(\ref{89-4}), (\ref{89-5}), (\ref{90-3}),
(\ref{7.3}), (\ref{92-1}) and (\ref{92-2}) show that the functions $A$ and $R$
depend on $U$, $V$ and $v_{r2}$ only through the combinations $-U + V$, $-U +
v_{r2}$ and $V - v_{r2}$. The shells given by Eqs.~(\ref{99-2}) and
(\ref{99-3}) are also shifted properly by (\ref{99-4}). Similar map
$\Phi^{\mathsf M}_t$ can be found for all cases if a gauge is chosen.

The structure that we need near $i^0$ is just the time measured by the
asymptotic observers; the range of this time is $(-\infty,\infty)$. We have to
define such a coordinate in each solution. The Schwarzschild time coordinate
$T_\infty$ in ${\mathcal M}_r$ for all cases A, A', B, B', C, C' seems to be a
good candidate. It must be defined in ${\mathcal M}_r$ by fixing it there with
respect to the already chosen coordinates $(U_r^+,V_r^+)$. This can be done as
follows:
\begin{equation}
  T_\infty := \frac{U_r^+ + V_r^+}{2}\ ;
\label{100-1}
\end{equation}

Above definition of the asymptotic coordinate $T_\infty$ in each solution is
formally analogous to choice of gauge; it is also similarly non
unique. However, all mathematically different choices of asymptotic time give
equivalent descriptions of the physical properties of the system.

Consider Case A and two solutions, one given by the values $M_m$, $v_{m2} -
v_{m1}$, $M_r$ and $v_{r2}$ of the parameters and the other by $M_m$, $v_{m2}
- v_{m1}$, $M_r$ and $v_{r2} + t$. Their ${\mathcal M}_r$-parts are mapped
isometrically on each other by $\Phi_t$ as follows
\[
  U_r^+ \mapsto U_r^+ + t\ ,\quad V_r^+ \mapsto V_r^+ + t\ .
\]
Hence,
\[
  T_\infty \mapsto T_\infty + t\ .
\]
This defines the action $\Phi^\infty_t$ of $\Phi_t$ at the infinity.

\subsection{Time reversal}
Consider the solution of case A determined by the values $M_m$, $v_{m2} -
v_{m1}$, $M_r$ and $v_{r2}$ of the parameters and the solution of case B
determined by $M'_m$, $u_{m2} - u_{m1}$, $M'_r$ and $u_{r2}$. Suppose that
$M'_m = M_m$, $M'_r = M_r$, $u_{m2} - u_{m1} = -(v_{m2} - v_{m1})$ and $u_{r2}
= -v_{r2}$. Then the two solutions are isometric and the corresponding map
inverts time orientation. This motivates the definition
\begin{equation}
  {\mathbf T}^P(M_m, v_{m2} - v_{m1},M_r,v_{r2}) = (M_m, -(v_{m2} -
  v_{m1}),M_r,-v_{r2}) 
\label{102-1}
\end{equation}
of the action ${\mathbf T}^P$ of the time reversal ${\mathbf T}$ on the
physical phase space of cases A and B. The inverse map is easily to calculate.

Similarly, for the case C, we define for the charts $(M_d,M_u,M_r,v_u)$
in the physical phase space of case C and $(M'_d,M'_u,M'_r,u'_d)$ in that of
case C':
\[
  {\mathbf T}^P(M_d,M_u,M_r,v_u) = (M_u,M_d,M_r,-v_u)\ .
\]
Consider the action of the time reversal on the background manifold $\mathsf
M$. One can choose the gauge in all cases so that always
\[
  {\mathbf T}^{\mathsf M}(U,V) = (-V,-U)\ .
\]
Finally, with our definition (\ref{100-1}) of the asymptotic time $T_\infty$,
we clearly have
\[
  {\mathbf T}^\infty(T_\infty) = -T_\infty\ .
\]

\subsection{Dilatation}
Consider again the formulae (\ref{89-4}), (\ref{89-5}), (\ref{90-3}),
(\ref{7.3}), (\ref{92-1}) and (\ref{92-2}) of case A. The transformation
\begin{alignat}{2}
  M_m & \mapsto e^\lambda M_m\ ,&\qquad M_r & \mapsto e^\lambda M_r \ ,
\label{63.1} \\
  v_{m2}-v_{m1} & \mapsto e^\lambda (v_{m2}-v_{m1})\ ,&\qquad v_{r2} & \mapsto
  e^\lambda v_{r2}\ ,
\label{63.2}
\end{alignat}
together with
\begin{equation}
  U \mapsto e^\lambda U\ ,\qquad V \mapsto e^\lambda V\ ,
\label{63.3}
\end{equation}
lead to $R \mapsto e^\lambda R$ and $A \mapsto A$, so that the metric is
rescaled as follows
\[
  -AdUdV + R^2d\Omega^2 \mapsto e^{2\lambda}(-AdUdV + R^2d\Omega^2)\ .
\]
The asymptotic time $T_\infty$ (see Eq.~(\ref{100-1})) is then transformed so
that $T_\infty \mapsto e^\lambda T_\infty$.

Analogous equations hold for all other cases. Eqs.~(\ref{63.1}) and
(\ref{63.2}) describe the action of the dilatation group $D_\lambda$ on the
space of the parameters and Eq.~(\ref{63.3}) that on the background manifold. 

The symmetry of the model under dilatation is, of course, due to the shell
matter being light-like.

\section*{Acknowledgments}
The authors are thankful for useful discussions by T.~Dray, W.~Israel, K.~V.~
Ku\-cha\v{r} and P.~Minkowski. The work has been supported by the Swiss
Nationalfonds and by the Tomalla Foundation, Zurich.


\begin{thebibliography}{99}
\bibitem{bergmann}P.~Bergmann, Rev. Mod. Phys. {\bf 33} (1961) 510.
\bibitem{torre}C.~Torre, Phys. Rev. D{\bf 46} (1993) R3231.
\bibitem{H-I}P.~H\'{a}j\'{\i}\v{c}ek and C.~Isham, J. Math. Phys. {\bf 37}
  (1996) 3522. 
\bibitem{haji} P.~H\'{a}j\'{\i}\v{c}ek, J. Math. Phys. {\bf 36} (1995) 4612,
  Class. Quant. Grav. {\bf 13} (1996) 1353. 
\bibitem{DW} B.~S.~DeWitt, Dynamical Theory of Groups and Fields, in {\it
    Relativity, Groups and Topology I}, eds. C.~DeWitt and B.~S.~DeWitt,
  Gordon and Breach, London (1964).  
\bibitem{A} A.~Ashtekar, {\it Asymptotic Quantization}, Bibliopolis, Napoli
  (1987)  
\bibitem{K}K.~V.~Kuchar, J. Math. Phys. {\bf 13} (1972) 768.
\bibitem{H-Kij}P.~H\'{a}j\'{\i}\v{c}ek and J.~Kijowski, Phys.~Rev.~D {\bf 61}
  (2000) 024037.
\bibitem{H-Kie}P.~H\'{a}j\'{\i}\v{c}ek and C.~Kiefer, Nucl.~Phys.~B {\bf 603}
  (2001) 491.
\bibitem{H}P.~H\'{a}j\'{\i}\v{c}ek, Nucl.~Phys.~B {\bf 603} (2001) 555.
\bibitem{B-H} J.~Bicak and P.~H\'{a}j\'{\i}\v{c}ek, Phys.~Rev. {\bf 56} (1997)
  4706.   
\bibitem{BI}C.~Barrabes and W.~Israel, Phys.~Rev.~D {\bf 43} (1991) 1129.
\bibitem{DtH}T.~Dray and G.~t'Hooft, Commun.~Math.~Phys.~{\bf 99} (1985) 613.
\bibitem{Red}I.~H.~Redmount, Progr.~Theor.~Phys.~{\bf 73} (1995) 1401.
\bibitem{witten} E.~Witten, Comm. Math. Phys. {\bf 80} (1981) 381.  
\bibitem{moncrief}J.~Marsden, {\em Lectures on Geometric Methods in
    Mathematical Physics,} CMBS Vol.~{\bf 37}, SIAM Philadelphia, 1981. 

\end{thebibliography}
\end{document}